\documentclass[a4,11pt]{article}
\title{The kinetics of homogeneous melting beyond the  \\
limit of superheating}
\author{D. Alf\`{e}$^{2,3,4,5}$, C. Cazorla$^{1}$ and 
M. J. Gillan$^{2,4,5}$ \\
$^1$Institut de Ci\`{e}ncia de Materials de Barcelona (ICMAB-CSIC), \\
Campus UAB, 08193 Bellaterra, Spain \\
$^2$Thomas Young Centre at UCL, London WC1E 6BT, UK \\
$^3$Department of Earth Sciences, UCL, London WC1E 6BT, UK \\
$^4$London Centre for Nanotechnology, UCL, London WC1H 0AH, UK \\
$^5$Department of Physics and Astronomy, UCL, London WC1E 6BT, UK}
\usepackage{a4wide}
\usepackage{graphicx}

\begin{document}
\maketitle
\abstract{Molecular dynamics simulation is used to study the
time-scales involved in the homogeneous melting of a superheated
crystal. The interaction model used is an embedded-atom model
for Fe developed in previous work, and the melting process is simulated
in the microcanonical $(N, V, E )$ ensemble. We study periodically
repeated systems containing from 96 to 7776 atoms, and the initial
system is always the perfect crystal without free surfaces or
other defects. For each chosen total energy $E$ and number of atoms $N$,
we perform several hundred statistically independent simulations,
with each simulation lasting for between 
$500$~ps and $10$~ns, in order to gather
statistics for the waiting time $\tau_{\rm w}$ before melting occurs.
We find that the probability distribution of $\tau_{\rm w}$ is roughly
exponential, and that the mean value $\langle \tau_{\rm w} \rangle$
depends strongly on the excess of the initial steady temperature of
the crystal above the superheating limit identified by other researchers.
The mean $\langle \tau_{\rm w} \rangle$ also depends strongly on
system size in a way that we have quantified. For very small systems
of $\sim 100$ atoms, we observe a persistent alternation between
the solid and liquid states, and we explain why this happens. Our results
allow us to draw conclusions about the reliability of the recently
proposed Z~method for determining the melting properties of simulated
materials, and to suggest ways of correcting for the errors of the method.}

\section{Introduction}
\label{sec:intro}

The supercooling of liquids below their thermodynamic freezing point
is familiar and easily observable, but the superheating of solids
above their melting point is much more difficult to study.  This is
because melting is usually initiated at surfaces (grain boundaries and
other defects may also initiate melting), so that superheating 
is generally possible only in
solids that have no surfaces~\cite{cahn86}. Melting
from the defect-free superheated state, sometimes called ``homogeneous
melting'', has been experimentally
observed~\cite{boyce85,rossouw85,dages86,evans86}, but there is still
rather little detailed understanding of the kinetics of the process.
Fortunately, computer simulation offers a rather straightforward way
of studying superheated defect-free solids, and this has led to a
recent resurgence of interest in the subject~\cite{jin01}. In addition to
the purely scientific interest, it has recently been shown that the
concept of the ``superheating limit'' leads naturally to a simulation
technique known as the ``Z~method'' that offers a new and potentially
useful way of determining the melting properties of simulated
materials~\cite{belonoshko06}.  In the present paper, we report new
simulation results on the kinetics of homogeneous melting which shed
light on the conditions needed for the Z~method to yield reliable
results.

The study of melting properties by computer simulation dates back over
50 years~\cite{alder57}. Two main approaches have become firmly
established over that period. The first relies on the separate
calculation of the free energies of the solid and liquid, and is based
on the fact that the two phases coexist in thermal equilibrium when
both the pressures and the chemical potentials in the solid and liquid
are equal~\cite{broughton87,foiles89,mei92}.  The second approach
consists of the explicit simulation of coexisting solid and liquid in
the same simulated system~\cite{morris94}. For any given interaction
model, careful application of the two approaches to large enough
systems should yield essentially identical results for the relation
between melting temperature $T_{\rm m}$ and the pressure $P$ on the
melting curve, as well as other properties, including the heat and
volume of fusion. The two approaches have been extensively used to
determine the melting properties of a wide variety of systems
interacting via classical potentials, including hard
spheres~\cite{alder57,choi91}, inverse-power~\cite{laird92,agrawal95} and
Lennard-Jones models~\cite{johnson93}, as well as more complex models such as
the Born-Mayer model of ionic liquids~\cite{anwar03}, a variety of
models for water~\cite{fernandez06}, and the embedded-atom model for
metals~\cite{daw93,belonoshko97}. In the past 15 years, there has been rapidly
increasing interest in the determination of melting properties using
{\em ab initio} molecular dynamics simulation (AIMD) based on
density-functional theory (DFT). Initially, the free-energy route was
used~\cite{sugino95,dewijs98,alfe99c}, with thermodynamic integration
employed to compute the difference of free energy between the {\em ab
 initio} system and an appropriately chosen reference system. The
coexistence approach has also been extensively used, mainly with
parameterised classical potentials tuned to data produced by DFT
simulations on the solid and liquid. However, there have also been
several studies in which direct {\em ab initio} simulations of
coexisting solid and liquid have been performed on systems of several
hundred atoms~\cite{alfe03,ogitsu03,bonev04,alfe05,alfe09,yoo09}.

The point of departure of the Z~method~\cite{belonoshko06} is 
an apparently simple question
about the superheating of a solid: If a solid is allowed to evolve
under the classical equations of motion at constant number of atoms $N$, volume
$V$ and internal energy $E$, what is the maximum energy $E_{\rm LS}$ it can 
have without eventually transforming completely into the liquid state? The 
proposed answer~\cite{belonoshko06} is
that it is the lowest energy on the given isochore within the field of
thermodynamic stability of the liquid~\cite{complete}.
This energy $E_{\rm LS}$ corresponds to a temperature $T_{\rm LS}$ 
representing the limit of superheating, above which the solid, evolving at
constant energy, will always melt. Since $E_{\rm LS}$ is
the lowest energy of the liquid on the isochore, it should be the energy 
of the liquid in coexistence with the solid, so that the energy $E_{\rm LS}$
of the liquid is associated with the melting temperature $T_{\rm m}$:
\begin{equation}
E^{\rm sol} ( V, T_{\rm LS} ) = E^{\rm liq} ( V, T_{\rm m} ) \; .
\label{eqn:Z_formula}
\end{equation}
The procedure for determing the point $( T_{\rm m} , P )$ on the melting curve
belonging to a specified liquid-state $V$ is then to perform a sequence
of $( N, V, E )$ m.d. simulations, monitoring $T$ and $P$ in each, the aim
being to locate the threshold $E_{\rm LS}$ (equivalently, the threshold
$T_{\rm LS}$), above which the transition to the liquid always occurs,
and below which it never occurs~\cite{belonoshko06}.

Implicit in these statements is an important question about
timescales, i.e. about the kinetics of homogeneous melting. The
original papers about the Z~method~\cite{belonoshko06} emphasise that
in order to be reasonably certain whether the initial $T$ is above or
below $T_{\rm LS}$, evolution must be allowed over a long enough time,
which may be on the order of ns, and the number of atoms $N$ must also
be large enough. (For simplicity, we assume here a single-component
system of atoms.) This naturally raises a number of important
questions, which we shall try to answer.  First, since homogeneous
melting in constant-$(N, V, E)$ dynamics appears to be a rare-event
process, we want to examine the probability distribution of waiting
times $\tau_{\rm w}$ before the transition occurs.  This means
repeating the simulation many times with the same $( N, V, E )$ but
with statistically independent initial conditions.  Second, we want to
study how the mean waiting time $\langle \tau_{\rm w} \rangle$ depends
on how far above $T_{\rm LS}$ the system is initiated. Third, we need
to determine the dependence of $\langle \tau_{\rm w} \rangle$ on the
system size $N$. Naturally, the numerical answers to these questions
will depend on the nature of the system and the number density $n = N
/ V$. Given the recent interest in using the Z~method for the
melting of metals~\cite{bouchet09}, particularly at high
pressures~\cite{belonoshko08,belonoshko09}, we
have decided to study the statistics of $\tau_{\rm w}$ using an
embedded atom model (EAM) for Fe, whose melting properties are already
well known from previous work~\cite{alfe02b}. We will also present Z-method
calculations on the melting of Fe using AIMD. We take a density
corresponding to the megabar pressures typical of the Earth's core.

In the next Section, we summarise the details of the interaction model
and the simulation procedures. Our results on the statistical
distribution of $\tau_{\rm w}$ and the dependence of
$\langle \tau_{\rm w} \rangle$ on initial and final temperatures and system
size from EAM and AIMD simulations are presented in 
Sec.~3. In the final Section, we discuss the
implications of our results for the understanding of homogeneous
melting and for the reliability of the Z~method, particularly in the
context of {\em ab initio} simulations.

\section{Techniques}
\label{sec:techniques}

The embedded-atom model (EAM) for Fe used in our simulations is the
one used as a reference system in our earlier work~\cite{alfe02b} on
the {\em ab initio} melting curve of hcp Fe. Essentially the same EAM
was also used in the very recent work of Belonoshko {\em et
  al.}~\cite{belonoshko09}, in which they used the Z method to study
the melting of Fe and an Fe/Si alloy. The model is actually a
modification of a much earlier EAM developed originally by
Belonoshko's group~\cite{belonoshko2000}. We recall that in an EAM
scheme the total potential energy $E_{\rm tot}$ is expressed as a sum
of atomic energies: $E_{\rm tot} = \sum_i E_i$, with the sum running
over the $N$ atoms in the system. Each term is a sum of two parts:
$E_i = E_i^{\rm rep} + F ( \rho_i )$.  Here, $E_i^{\rm rep}$ consists
of a sum of repulsive inverse-power pair potentials: $E_i^{\rm rep} =
{\sum_j}^\prime \epsilon ( a / r_{i j} )^n$, where $r_{i j}$ is the
distance between atoms $i$ and $j$, and the term $i = j$ is
excluded. $F ( \rho_i )$ is an ``embedding function'' which describes
the metallic bonding. It has the form $F ( \rho_i ) = - \epsilon C
\rho_i^{1/2}$, with $\rho_i = {\sum_j}^\prime ( a / r_{i j} )^m$. The
values of $a$ and $m$ are those in the original Belonoshko
model~\cite{belonoshko2000}, while in Ref.~\cite{alfe02b} we showed how
all the other parameters could be optimised by minimising the
fluctuations of the difference between the EAM and {\em ab initio}
energies in simulations of the liquid and the high-temperature
solid. The numerical values of the parameters are: $a = 3.4714$~\AA,
$m = 4.788$, $\epsilon = 0.1662$~eV, $n = 5.93$, $C = 16.55$. We apply
the spatial cut-off $r_c = 5.5$~\AA, so that terms in both $E_i^{\rm
  rep}$ and $\rho_i$ for which $r_{i j} > r_c$ are set to zero, with
the usual cutting and shifting used to ensure continuity.

Following previous work~\cite{alfe99a,alfe01,alfe02}, we focus here on
the melting properties of Fe in the high-pressure region that is
important for understanding the solid inner core and the liquid outer
core of the Earth. Specifically, we confine ourselves to pressures $P
\simeq 330$~GPa, which is the pressure at the boundary between the
inner and outer core~\cite{poirier91}. From extensive simulations with
our EAM on large systems containing solid and liquid in stable
coexistence~\cite{alfe02b}, we know that its melting temperature at $P =
323$~GPa is $6200 \pm 100$~K. We have recently refined these
coexistence simulations so as to reduce the statistical errors,
finding that a more accurate value of $T_{\rm m}$ at this pressure is
$6215$~K.

All the simulations to be presented were performed at the same density
corresponding to a volume per atom $V / N = 7.139$~\AA$^3$, which gives
pressures in the region of interest. In every simulation, we start
from the perfect hcp crystal, with all atoms on their regular lattice
sites, and we assign random velocities drawn from a Maxwellian
distribution, the velocities then being shifted and scaled so that the
total momentum is zero and the kinetic energy per atom $K / N$ has a
value corresponding exactly to a chosen initial temperature $T_{\rm i}
= 2 K / 3 k_{\rm B} N$.  Verlet's algorithm~\cite{verlet67} 
was used with a time-step
of $1$~fs, which ensures conservation of total energy with a drift
of typically no more than $\sim 10$~K over times of several ns.

We shall present simulations on systems containing $N = 96$, $150$,
$392$, $972$ and $7776$ atoms. For each $N$ and for each initial
temperature $T_{\rm i}$, we have performed several hundred m.d.
simulations of at least $500$~ps (in some cases we have continued the
simulations for over $10$~ns), in order to gather statistical
information about the melting process. For the larger systems, the
simulations were run on large parallel computers, with each individual
simulation running on $24$ cores, and with typically $350$ such
simulations running simultaneously. This mode of operation makes it
possible, for example, to run an overall total of $\sim 1$~$\mu$s of
m.d. on the $7776$-atom system in only a few hours of wall-clock
time. For the small systems, we ran the calculations on single
processors using local facilities.

The {\em ab-initio} simulations were run with exactly the same
technical details as described in~\cite{alfe99a,alfe01,alfe02}, using
the {\sc vasp} code~\cite{kresse96} with the projector augmented wave
method~\cite{blochl94,kresse99} and an efficient extrapolation of the
charge density~\cite{alfe99b}.

\section{Results}
\label{sec:results}

\begin{figure}
\centering
{\includegraphics[width=0.85\linewidth]{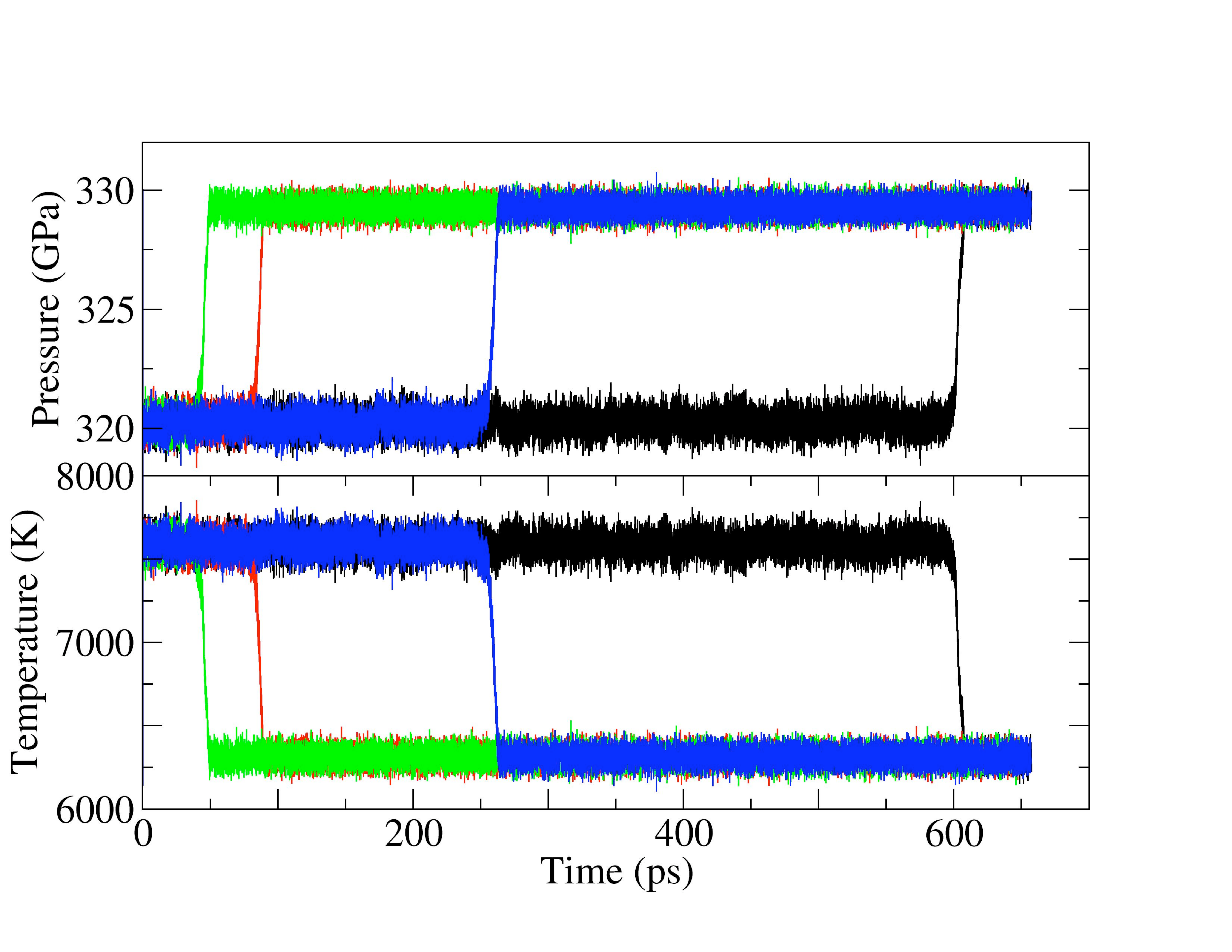}}
\vspace{0.1cm}
\caption{Time-dependent temperature and pressure in four independent
simulation runs, showing homogeneous melting from
superheated hcp solid Fe in a system of 7776 atoms. All four simulations
were initiated from perfect crystal positions, with initial random
velocities corresponding to the same temperature $T_{\rm m} = 15600$~K,
the mean quasi-steady temperatures of the superheated solid and the
final liquid being $T_{\rm sol} = 7590$~K and $T_{\rm liq} = 6315$~K.}
\label{fig:tdep_TP_7776}
\end{figure}

We start by showing some examples of homogeneous melting from our
simulations on the system of $7776$ atoms. Fig.~\ref{fig:tdep_TP_7776}
displays the time-dependent temperature and pressure in four
simulations that were all initiated from the perfect hcp crystal with
exactly the same kinetic energy corresponding to the temperature
$T_{\rm i} = 15600$~K, but with statistically independent random
velocities. As expected from equipartition, $T$ drops rapidly to about
half its initial value (this rapid drop is not shown in the Figure),
and it then fluctuates about a quasi-steady value $T_{\rm sol} =
7590$~K for many ps, before it drops again over a rather short period
of $\sim 8$~ps, and then fluctuates again about a lower steady value
$T_{\rm liq} = 6315$~K. The second drop is due to melting, as we have
verified by monitoring the self-diffusion of atoms via the
time-dependent mean-square displacement $\langle \Delta r ( t )^2
\rangle$. The appearance of atomic disorder throughout the system when
the system melts is also easy to observe in movies prepared from the
coordinate files. Melting is accompanied by an increase of $P$ by
$\sim 10$~GPa, which occurs over the same rather short interval as the
drop in $T$. These effects are familiar from many previous reports on
the Z~method~\cite{belonoshko06,belonoshko08,belonoshko09}: the drop
in $T$ is due to the latent heat of fusion, and the increase of $P$ is
associated with the volume increase that would occur on melting if the
pressure were held constant. We note that the waiting times $\tau_{\rm
  w}$ that elapse before melting are different in the four examples
shown. This is what we expect of a rare-event process, and is
consistent with the statements in earlier
reports~\cite{belonoshko06,belonoshko08,belonoshko09} that the time at
which the melting transition occurs is not correlated with the details
of the initial conditions. We find that the final mean temperature
$T_{\rm liq}$ and pressure $P_{\rm liq}$ of the liquid are the same in
all the examples. This is expected, because in every case the system
settles into exactly the same thermodynamic state of the liquid.  The
temperature $T_{\rm liq}$ is somewhat above the melting temperature
$T_{\rm m}$ at pressure $P_{\rm liq}$, as expected because the system
was initiated above the limit of superheating.

These observations naturally raise the question of the statistical
distribution of waiting times $\tau_{\rm w}$. To investigate this,
we have to repeat the simulations many times, starting always from
the perfect lattice with exactly the same initial kinetic energy, but
independent random velocities. To make the question well posed, we need
a definition of $\tau_{\rm w}$. We note from 
Fig.~\ref{fig:tdep_TP_7776} that the fluctuations
of $T$ about its mean value in the quasi-steady state of the solid before
melting are much smaller than the drop of $T$ during the melting process.
We therefore define $\tau_{\rm w}$ to be the elapsed time from the start
of the simulation to the instant when $T$ is mid-way between the mean
quasi-steady temperature $T_{\rm sol}$ of the solid and the mean 
final temperature $T_{\rm liq}$ of the liquid.

\begin{figure}
\centering
{\includegraphics[width=0.85\linewidth]{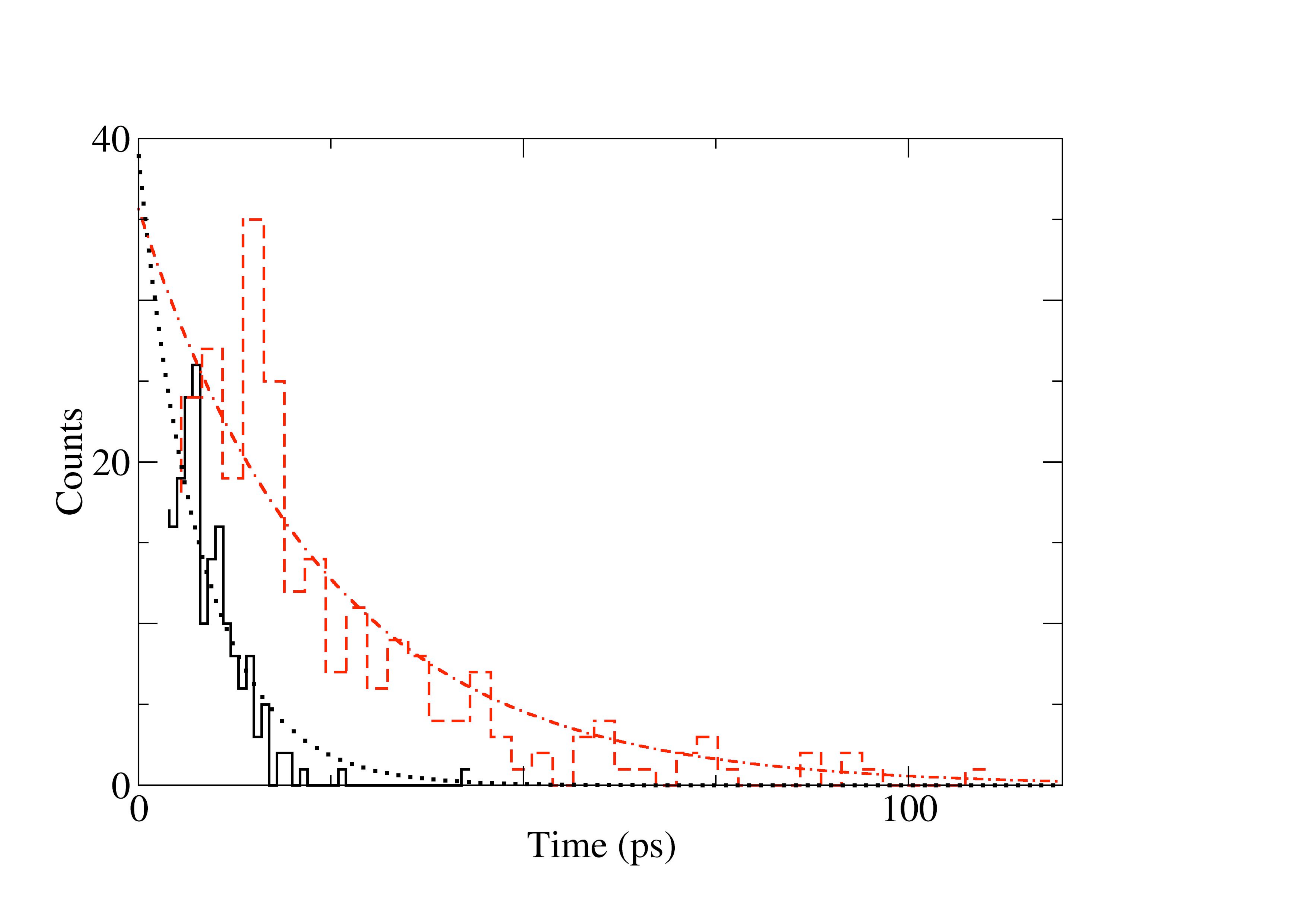}}
\vspace{0.1cm}
\caption{Histograms of waiting times $\tau_{\rm w}$ before the
transition to liquid constructed from repeated simulations at
two initial temperatures for the system of 7776 atoms. Histograms
shown by dashed (red) and solid (black) lines result from initial
temperatures of $T_{\rm i} = 15800$ and $16000$~K respectively,
the quasi-steady solid and liquid temperatures in the two cases
being $T_{\rm sol} = 7640$ and $7740$~K and $T_{\rm liq} = 6410$
and $6505$~K. Dashed and dotted curves show exponential functions
fitted to histograms (see text).}
\label{fig:hist_wait_7776}
\end{figure}

For the $7776$-atom system with $T_{\rm i} \simeq 15800$~K, we 
repeated the simulations 350 times, with each run having a duration
of 650~ps. We found that it melted in all cases, and we  
accumulated the histogram of $\tau_{\rm w}$ 
shown in Fig.~\ref{fig:hist_wait_7776}.
We see that after a short incubation time of no more than $\sim 20$~ps,
the probability distribution of $\tau_{\rm w}$ decays in a 
quasi-exponential way. This is what we expect if melting is a random
process having short memory time with a constant probability per unit
time $1 / \tau_0$ of occurring, given that it has not already occurred.
In this case, the probability distribution of waiting times 
$p ( \tau_{\rm w} )$ would have the form
$\tau_0^{-1} \exp ( - \tau_{\rm w} / \tau_0 )$ and the mean waiting time would
be $\langle \tau_{\rm w} \rangle = \tau_0$.
We show in Fig~\ref{fig:hist_wait_7776} a fit of 
the exponential function to the histogram.
The value of $\tau_0$ given by this fit is
$\tau_0 = 24.1$~ps, which agrees well with the value
$\langle \tau_{\rm w} \rangle = 24.7$~ps computed directly from
the sample of 350 values of $\tau_{\rm w}$. We have checked that
the final mean $T_{\rm liq}$ and $P_{\rm liq}$ of the liquid are the same
in all cases, within statistical error, having the numerical values
$T_{\rm liq} = 6410 \pm 5$~K and $P_{\rm l} = 330.4 \pm 0.3$~GPa.

For comparison, we show the results for $p ( \tau_{\rm w} )$ when
we do exactly the same thing for the $7776$-atom system, but now
with a higher initial $T_{\rm i}$ of 16000~K, the number of independent
simulations in this case also being $350$. As expected, melting occurs
on average more rapidly with this $T_{\rm i}$, the values of $\tau_0$
from the exponential fit and from the directly computed
$\langle \tau_{\rm w} \rangle$ being $8.3$~ps and $9.4$~ps. The mean
temperatures of the solid and the liquid in this case are
are $T_{\rm sol} = 7750$~K and $T_{\rm liq} = 6510$~K, which, as expected,
are higher than the values found with the lower $T_{\rm i}$.

It is clear from these observations that the mean waiting time depends
on the extent to which the temperature $T_{\rm sol}$ exceeds the limit
of superheating $T_{\rm LS}$. To study this further, we have repeated
the simulations with $T_{\rm i}$ values of $16000$, $15800$, $15600$,
$15400$ and $15200$~K.  At the lowest of these $T_{\rm i}$, melting
was not seen in any of the simulations, even though we repeated them
350 times with statistically independent initial velocities, the
duration of the simulation being $710$~ps in every case. At $T_{\rm i}
= 15400$~K and $T_{\rm i} = 15600$~K , melting was observed within
$660$~ps in only 14 and 283 out of 350 simulations, respectively. At
all the other $T_{\rm i}$ values, melting occurred in all 
cases, and we were able to construct essentially complete
histograms; the values of $\langle \tau_{\rm w} \rangle$ and the value
of $\tau_0$ obtained by fitting $p ( \tau_{\rm w} ) = \tau_0^{-1} \exp
( - t / \tau_0 )$ to the histogram agreed closely. 

\begin{figure}
\centering
{\includegraphics[width=0.85\linewidth]{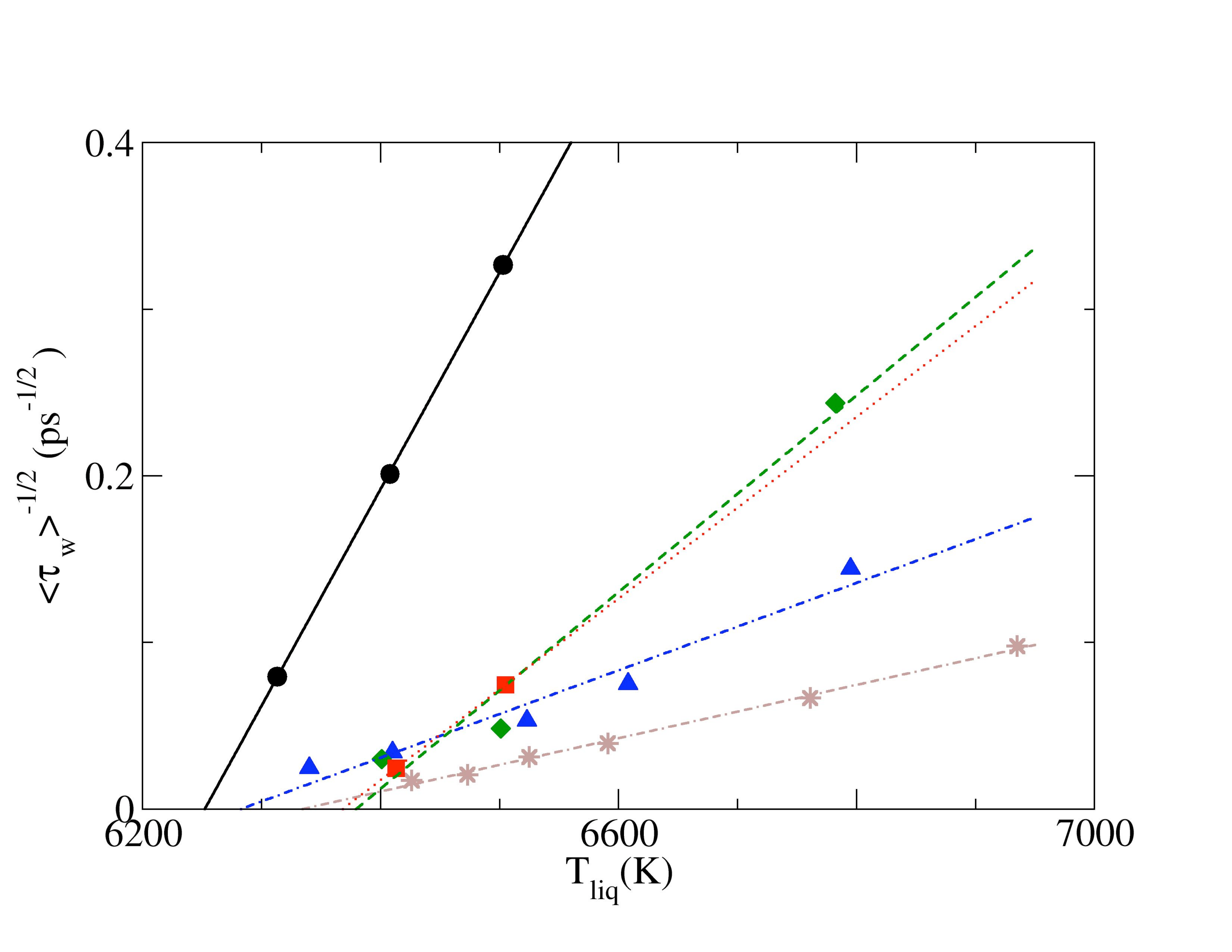}}
\vspace{0.1cm}
\caption{Dependence of mean waiting time $\langle \tau_{\rm w}
  \rangle$ on final liquid temperature $T_{\rm liq}$ for systems of $N
  = 7776$ (black circles), $976$ (red squares), $392$ (green
  diamonds), $150$ (blue triangles) and $96$ (brown stars)
  atoms. Quantity plotted is $\langle \tau_{\rm w} \rangle^{-1/2}$ as
  function of $T_{\rm liq}$.  Straight lines are linear least-squares
  fits to the data for each $N$ value.}
\label{fig:mwait_Tliq}
\end{figure}

The final mean $T_{\rm liq}$ of the liquid is a monotonically
increasing function of $T_{\rm i}$, and it is convenient
to examine the dependence of $\langle \tau_{\rm w} \rangle$ on $T_{\rm liq}$.
We have found that it is helpful to plot the quantity 
$\langle \tau_{\rm w} \rangle^{-1/2}$ against $T_{\rm liq}$, as shown
in Fig.~\ref{fig:mwait_Tliq}. 
The points fall roughly on a straight line, and the
indication is that $\langle \tau_{\rm w} \rangle^{-1/2} \rightarrow 0$
(i.e. $\langle \tau_{\rm w} \rangle \rightarrow \infty$) at
a characteristic temperature. At the same time, $P_{\rm liq}$ also tends to
a limiting value. We identify the characteristic temperature
as the melting temperature $T_{\rm m}$ at the pressure $P_{\rm liq}$,
because $T_{\rm m}$ is the lowest possible final mean temperature of
the liquid, namely the temperature found when $T_{\rm sol} = T_{\rm LS}$.
The value of $T_{\rm m}$ obtained by this extrapolation is 6260~K,
the extrapolated pressure being $328$~GPa. These results agree
very well with the value $T_{\rm m} = 6215$~K from explicit
coexistence simulations at the pressure 
$P = 323$~GPa (see Sec.~\ref{sec:techniques}).

All the results presented so far are for the large system of $7776$ atoms.
We have performed simulations of essentially the same kind for systems
of $N = 972$, 392, 150 and 96 atoms, in each case initiating the simulations
at sequence of initial temperatures $T_{\rm i}$, repeating the
simulations at each $( N, T_{\rm i} )$ a few hundred times, determining
the liquid-state $T_{\rm liq}$ values for the cases where melting occurs,
and extracting the $\langle \tau_{\rm w} \rangle$ values.
The plots of $\langle \tau_{\rm w} \rangle^{-1/2}$ against $T_{\rm liq}$
for all the system sizes are displayed in 
Fig.~\ref{fig:mwait_Tliq}. The results appear
to be very coherent: for each $N$, the $\langle \tau_{\rm w} \rangle^{-1/2}$
points fall reasonably well on a straight line, and the Figure shows
the linear least-square fits. These linear fits extrapolate to give
$T_{\rm m}$ values that agree for the different system sizes to
within $\sim 100$~K, i.e. to within $\sim 2$~\%. This 
agreement suggests that the Z~method
can be a robust way of obtaining $T_{\rm m}$ values close to the
thermodynamic limit for systems that would be much too small 
for explicit coexistence simulations, provided very long simulations
are performed and provided one extrapolates to the 
$\langle \tau_{\rm w} \rangle \rightarrow \infty$ limit.

\begin{figure}
\centering
{\includegraphics[width=0.85\linewidth]{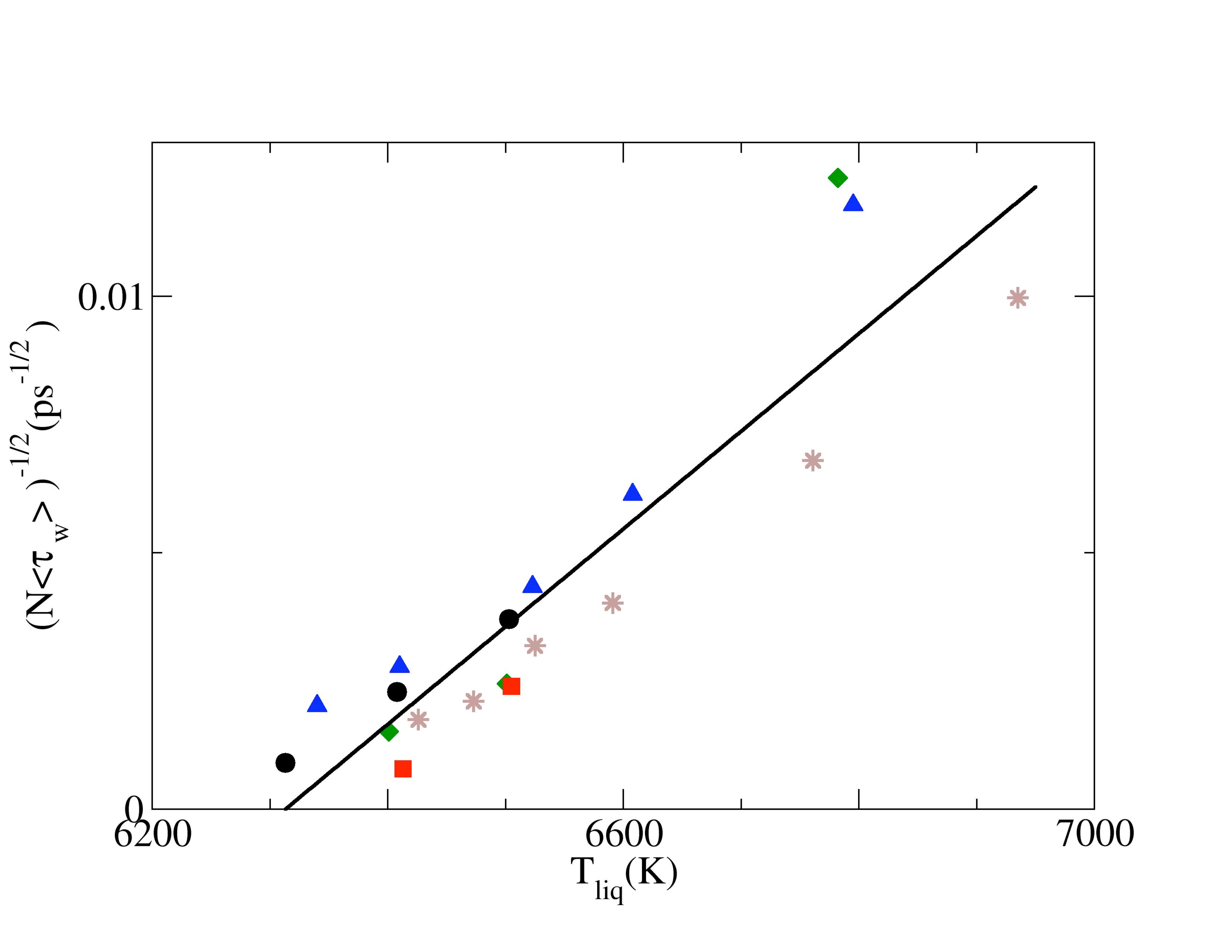}}
\vspace{0.1cm}
\caption{Scaling of mean waiting times $\langle \tau_{\rm w} \rangle$
with system size specified by number of atoms $N$. Quantity plotted is
$( N \langle \tau_{\rm w} \rangle )^{-1/2}$ as function of final
liquid temperature $T_{\rm liq}$.}
\label{fig:N_mwait_Tliq}
\end{figure}

The plots of Fig.~\ref{fig:mwait_Tliq} show 
that for the given density $n$ the mean
waiting times $\langle \tau_{\rm w} \rangle$ for different degrees
of superheating and different system sizes can all be roughly
represented by the formula 
$\langle \tau_{\rm w} \rangle = A / ( T_{\rm liq} - T_{\rm m} )^2$,
where $T_{\rm m}$ is independent of $T_{\rm liq}$ and $N$ but $A$
depends on $N$. They also show that $A$ increases with decreasing $N$.
More extensive results would be needed to make precise statements
about this $N$ dependence, but we find that the inverse proportionality
$A = B / N$ fits the results quite well.
As evidence for this, we display in Fig.~\ref{fig:N_mwait_Tliq}
a plot of $( N \langle \tau_{\rm w} \rangle )^{-1/2}$ against
$T_{\rm liq}$, showing that all our data are quite well reproduced by the
formula:
\begin{equation}
( N \langle \tau_{\rm w} \rangle )^{-1/2} =
C ( T_{\rm liq} - T_{\rm m} ) \; ,
\label{eqn:N_T_scaling}
\end{equation}
where $T_{\rm m} = 6315$~K and $C \equiv B^{-1/2}$ has the value 
$1.9 \times 10^{-5}$~ps$^{-1/2}$~K$^{-1}$.

\begin{figure}
\centering
{\includegraphics[width=0.85\linewidth]{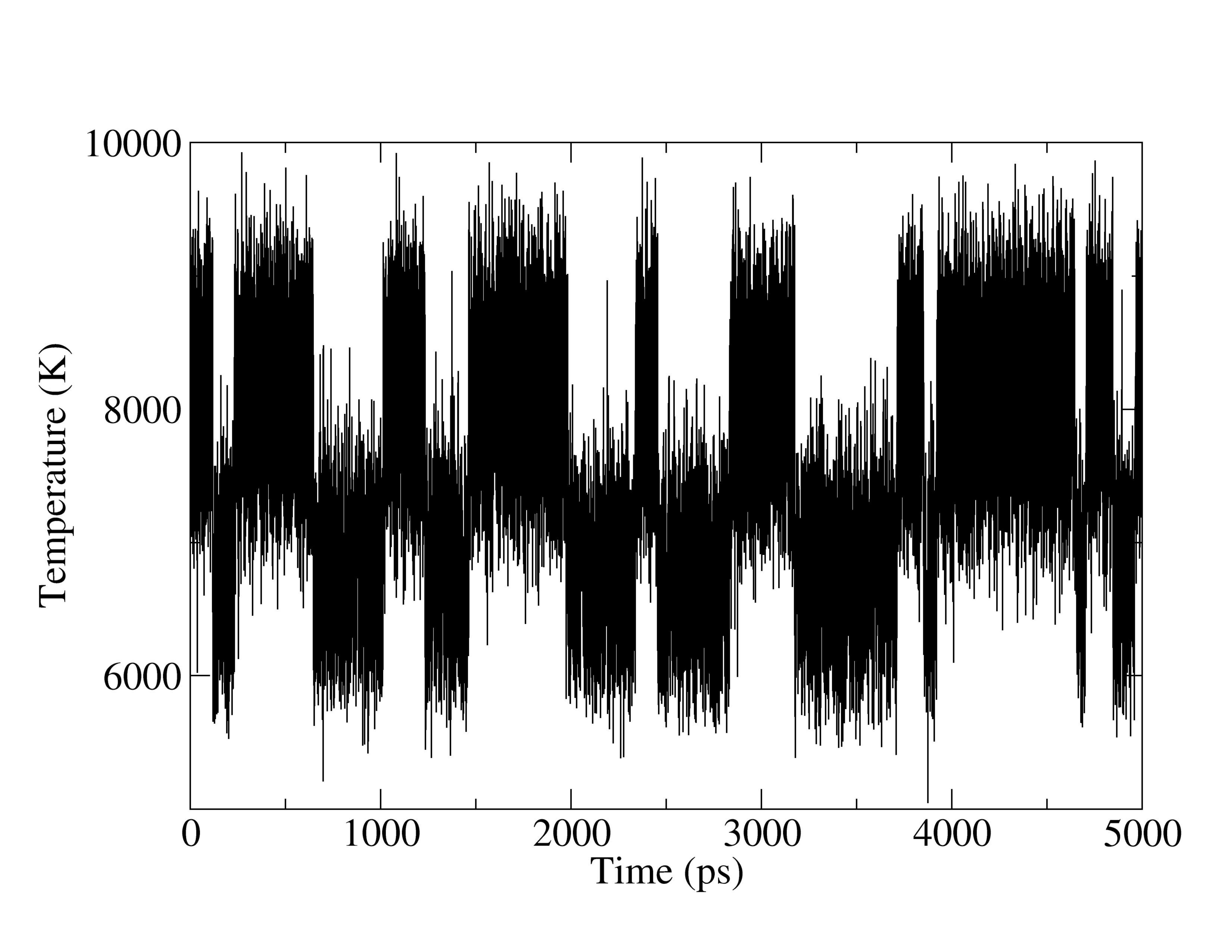}}
\vspace{0.1cm}
\caption{Alternation between solid and liquid states:
temperature as function of time in one of the constant-energy m.d. 
simulations on the system
of $96$ atoms, with total energy such that mean liquid-state
temperature is $T_{\rm liq} = 6760$~K, showing alternation between
mean temperatures $T_{\rm sol}$ and $T_{\rm liq}$.}
\label{fig:sol_liq_alt_96}
\end{figure}

\begin{figure}
\centering
{\includegraphics[width=0.85\linewidth]{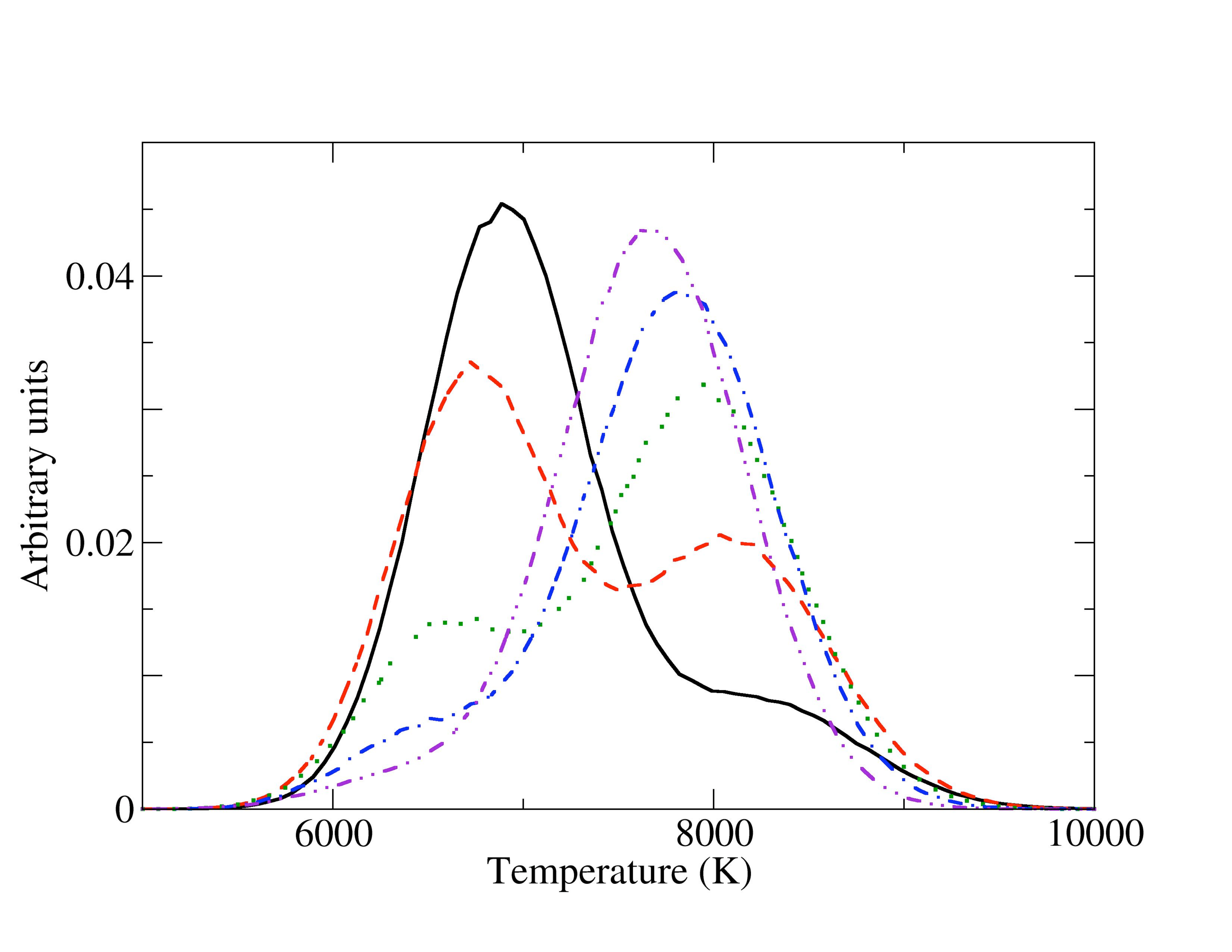}}
\vspace{0.1cm}
\caption{Histograms of temperature distribution at different
constant total energies $E$ in the system of $96$ atoms. The 
histogram at each $E$ was
obtained by sampling over typically 128 simulations, each having
a typical duration of 5~ns. Instead of giving $E$ directly,
we specify each histogram by the liquid-state temperature $T_{\rm liq}$.
Histograms shown by solid (black), dashed (red), dotted (green),
chain (blue) and dotted-chain (black) lines are for 
$T_{\rm liq} = 6935$, 6760, 6590, 6565 and 6473~K.}
\label{fig:hist_T_96}
\end{figure}

\begin{figure}
\centering
{\includegraphics[width=0.85\linewidth]{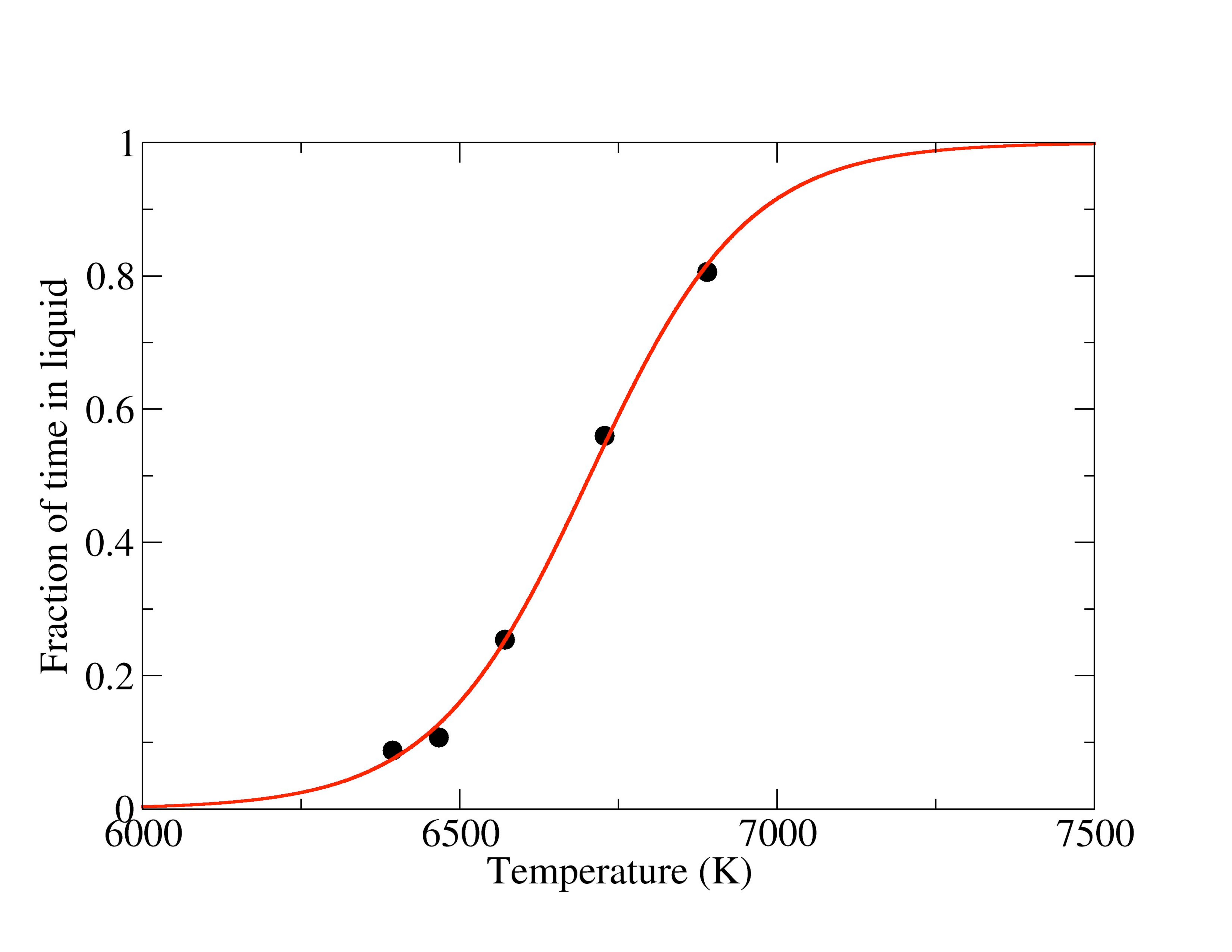}}
\vspace{0.1cm}
\caption{Fraction of time spent by the system in the liquid state
for different liquid-state temperatures $T_{\rm liq}$ in simulations
of 96-atom system.}
\label{fig:frac_liq_96}
\end{figure}

For the system of $96$ atoms, we observe a significant and interesting
effect, which sheds further light on the kinetics of homogeneous melting.
After the system has made the transition from superheated solid to liquid,
it remains in the liquid state only for a finite time, before reverting back
to the solid state. In fact, if the simulation is continued long enough,
we observe a continual alternation between the solid and liquid states.
An example of this behaviour is shown in 
Fig.~\ref{fig:sol_liq_alt_96}, where we see that the
temperature alternates between solid-like and liquid-like values $T_{\rm sol}$
and $T_{\rm liq}$. This effect becomes very clear if we construct a
histogram of temperature $T$ obtained by sampling over the course of many
simulations, all having exactly the same total energy $E$, and hence the
same liquid temperature $T_{\rm liq}$. The temperature histograms for 
the $96$-atom system for different $T_{\rm liq}$ values 
are shown in Fig.~\ref{fig:hist_T_96}.
We see that in each case $T$ has a bimodal distribution, being the
superposition of the quasi-Gaussian distributions that would be found if
the system were wholly in the solid state or wholly in the liquid state.
In fact, we can fit the histograms very well by a superposition of
Gaussians, and the relative weights of the two Gaussians for a given
$T_{\rm liq}$ tell us the relative amounts of time spent by the system
in the solid and liquid states. We display the fraction of time
$\alpha_{\rm liq}$ spent in the liquid state as a function of $T_{\rm liq}$
in Fig.~\ref{fig:frac_liq_96}.

At first sight, the ceaseless alternation between solid and liquid might
seem surprising, because it implies that homogeneous melting from
the superheated solid is not the irreversible process that one might
expect. However, in Sec.~\ref{sec:discussion} we will 
point out why this alternation is
required by the principles of statistical mechanics, we give
a simple formula that explains why $\alpha_{\rm liq}$ depends on 
$T_{\rm liq}$ as shown in Fig.~\ref{fig:frac_liq_96}, 
and we discuss whether the alternation
should also be seen in larger systems.

There has been considerable interest in using the Z-method with AIMD
simulation (we refer to this as AIMD-Z) to obtain melting curves,
particularly for metals~\cite{belonoshko08,belonoshko09}. To test the
practical operation of AIMD-Z, we have performed our own calculations
on the high-$P$ melting of hcp~Fe, on which there is already very
extensive previous work based on both free-energy and
explicit-coexistence methods, including a recent study of AIMD
coexistence on systems of $N = 980$ atoms~\cite{alfe09}. Before
presenting our AIMD-Z results on this problem, it is useful to
consider the errors that can be expected. Our AIMD-Z simulations were
performed on systems of 150 atoms with duration of $\sim
50$~ps. Clearly, with a run of this duration, initiated above $T_{\rm
  LS}$, we are unlikely to observe homogeneous melting unless $\langle
\tau_{\rm w} \rangle$ is $\sim 50$~ps or less.  From the formula given
in eqn~(\ref{eqn:N_T_scaling}), we estimate that this will yield
$T_{\rm liq} - T_{\rm m} \simeq 600$~K, and this is the error we may
make if $T_{\rm m}$ is estimated as the lowest $T_{\rm liq}$ for which
homogeneous melting is observed.

\begin{figure}
\centering
{\includegraphics[width=0.85\linewidth]{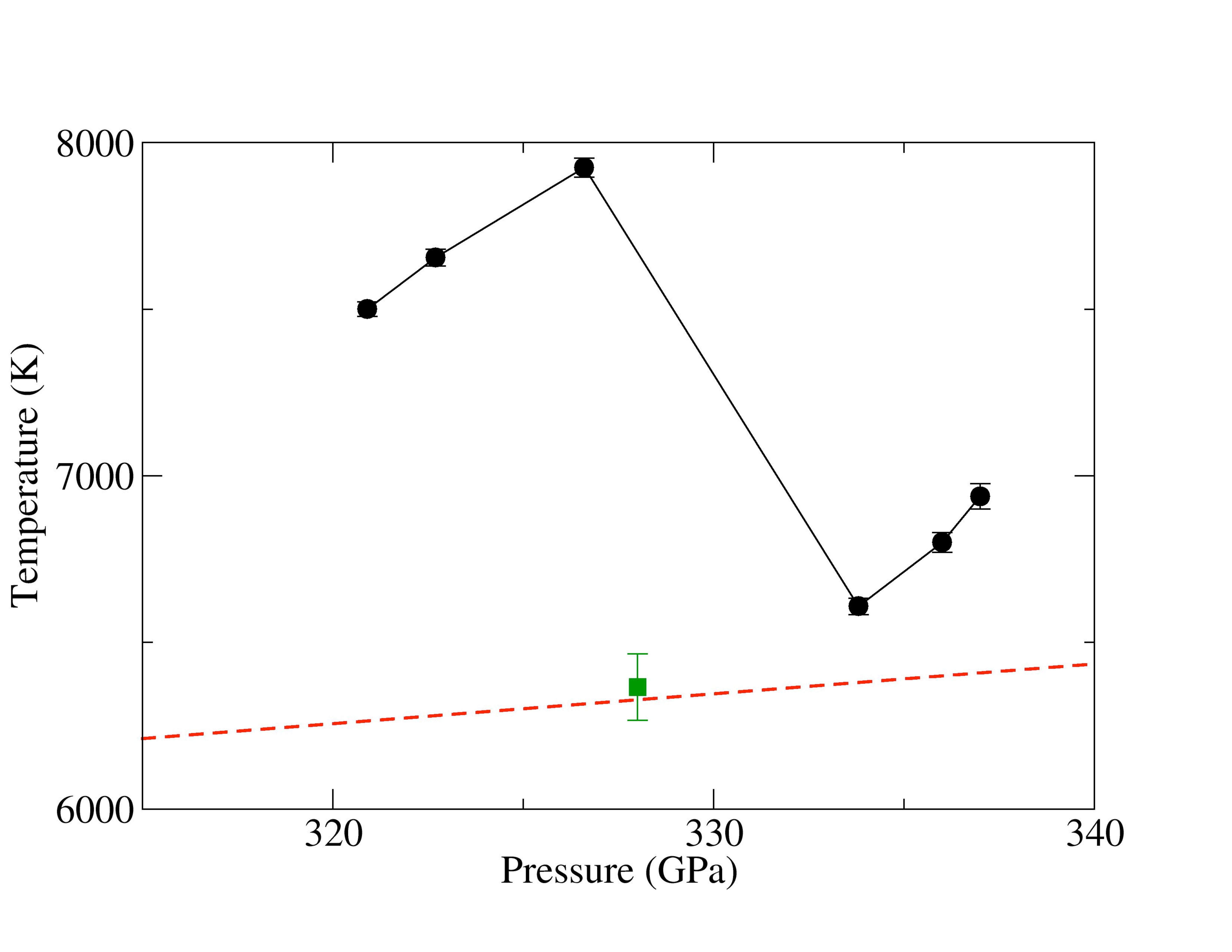}}
\vspace{0.1cm}
\caption{Z plot from a sequence of constant-energy AIMD 
simulations on system of 150
atoms, duration of simulations being $50$~ps. Black filled circles with
error bars show final mean temperature and pressure, with upper-left
branch corresponding to energies for which the system remains solid, and
lower right branch to energies for which homogeneous melting occurs. Dashed
(red) line shows the {\em ab initio} melting curve obtained in earlier work 
using the free energy technique, and green filled square with error bar
shows point on {\em ab initio} melting curve obtained with {\em ab initio}
m.d. simulations on a system of 980 atoms containing coexisting solid ald
liquid.}
\label{fig:Z_AIMD}
\end{figure}

We present in Fig.~\ref{fig:Z_AIMD} our AIMD-Z results for the melting
of hcp Fe with $N = 150$ using $50$-ps simulations. The Figure also
shows the melting curve obtained many years ago with exactly the same
AIMD techniques but based on free-energy
calculations~\cite{alfe01,alfe02,alfe99b}, as well as a point on the
melting curve from AIMD coexistence using 980 atoms~\cite{alfe09}. As
expected, AIMD-Z overestimates $T_{\rm m}$, and the amount of the
overestimate $T_{\rm liq} - T_{\rm m}$ is similar to our prediction
from eqn~(\ref{eqn:N_T_scaling}).


\section{Discussion and conclusions}
\label{sec:discussion}

Our investigation has yielded several simple but important insights
into the kinetics of homogeneous melting under $( N, V, E )$
conditions.  First, we have confirmed the existence of a ``limit of
superheating'' $T_{\rm LS}$ proposed in previous
work~\cite{belonoshko06}, and we have shown
that for a given ``excess'' quasi-steady temperature $T_{\rm sol} -
T_{\rm LS}$ of the initial superheated solid, and for a given number
of atoms $N$, there is a fairly well defined probability per unit time
(reciprocal of mean waiting time $\langle \tau_{\rm w} \rangle$) of
making the melting transition.  For given $N$, $\langle \tau_{\rm w}
\rangle$ appears to scale roughly as $A / ( T_{\rm liq} - T_{\rm m}
)^2$ with the excess of the final liquid temperature $T_{\rm liq}$
above the true thermodynamic melting temperature $T_{\rm m}$ associated
with the specified liquid density.  The coefficient $A$ appears to be
roughly proportional to $1 / N$.  These rather simple findings are
clearly interesting, and it is natural to ask whether they will hold
true for materials other than the particular transition metal studied
here. At present, we have no way of answering this question, and there
is now a clear need to extend the investigation to materials of other
kinds.

Another natural question concerns the relation between homogeneous melting
and the kinds of metastable behaviour well known in, for example,
supersaturated vapours or supercooled liquids. Of course, theories
of such phenomena have an extremely long history, the point of
reference often being ``classical nucleation 
theory'' (CNT)~\cite{abraham74}. In CNT,
the waiting time for the transition to the thermal equilibrium state
(condensation, freezing,...) is governed by the time needed to form
a ``critical nucleus''; the associated free-energy barrier results
from a competition between
the lowering of bulk free energy resulting from the transition and
the free energy increase due to the formation of interfaces (liquid-vapour,
solid-liquid,...) during the transition. 
The metastable behaviour of a superheated solid
that is only slightly beyond the thermal-equilibrium stability field
of the solid should also be describable by an appropriate CNT.
However, it is possible that the formation of a critical nucleus
described by such a CNT has little to do with the homogeneous melting
observed in previous simulations~\cite{jin01} and 
in the ones presented here, for
two reasons. First, CNT theories and other theories of nucleation
would not predict a ``superheating limit''. Instead, they would predict
a mean waiting time for nucleation that decreases continuously as we move
further into the field of thermodynamic instability. Second, the
transition associated with classical nucleation will generally lead
to a final state in which both phases are present, rather than the
single (liquid) phase seen in simulations of homogeneous melting.

The persistent alternation between solid and liquid states that we observe
for the very small $96$-atom system is relevant here. To understand why
this happens, and to know when we should expect to see it, we recall the
ergodic principle, which is generally accepted to hold for condensed-matter
systems. This states that the trajectory produced by $( N, V, E )$
m.d. starting from any phase-space point (set of positions and momenta)
having specified total energy $E$ will, if continued long enough, pass
arbitrarily close to an arbitrarily chosen phase-space point of the same $E$.
The configurations we are concerned with here are either solid-like or
liquid-like: in none of the simulations we have performed do we see solid
and liqid simultaneously present, so that the configurations are either one
or the other, except for the very small fraction of configurations
that occur during the transitions from solid to liquid or vice versa.
Now suppose that, starting from the superheated solid, the system has
homogeneously melted and become liquid. Then the ergodic principle
tells us that the given trajectory, if continued long enough, will
eventually re-enter regions of solid-like configurations, so that it
will re-freeze. Indeed, the trajectory will densely cover the 
entire constant-$E$ manifold, and will spend well defined fractions
$\alpha_{\rm liq}$ and $\alpha_{\rm sol} \equiv 1 - \alpha_{\rm liq}$ 
in the liquid and solid states. This is exactly what we have seen, and
Fig.~\ref{fig:frac_liq_96} displays the 
value of $\alpha_{\rm liq}$ as a function of $T_{\rm liq}$.

It is straightforward to confirm that this explanation is correct.
The fractions $\alpha_{\rm liq}$ and $\alpha_{\rm sol}$ are proportional
to the numbers $W_{\rm liq}$ and $W_{\rm sol}$ of liquid-like and
solid-like microstates on the given constant-$E$ manifold. But these
are related to the entropies $S_{\rm liq}$ and $S_{\rm sol}$
of the corresponding macrostates:
$S_{\rm sol, liq} = k_{\rm B} \ln W_{\rm sol, liq}$. Hence, we have:
\begin{equation}
\alpha_{\rm liq} / \alpha_{\rm sol} = 
\alpha_{\rm liq} / ( 1 - \alpha_{\rm liq} ) =
\exp \left( \left( S_{\rm liq} - S_{\rm sol} \right) / k_{\rm B} \right) \; .
\label{eqn:alpha_liq_sol}
\end{equation}
This means that if we choose $E = E_0$ so that the system spends equal
amounts of its time in solid and liquid states, then the entropies are
equal. Let the temperatures in this situation be $T_{\rm sol}^0$ and
$T_{\rm liq}^0$. If we go to a nearby energy $E = E_0 + \delta E$,
then $\alpha_{\rm liq}$ and $\alpha_{\rm sol}$ change, because:
\begin{equation}
S_{\rm liq} - S_{\rm sol} = \left(
\left( \frac{\partial S_{\rm liq}}{\partial E} \right)_V -
\left( \frac{\partial S_{\rm sol}}{\partial E} \right)_V 
\right) \delta E \; .
\label{eqn:S_liq_sol}
\end{equation}
But $( \partial S / \partial E )_V = 1 / T$, and $\delta E$ can be
expressed as $\delta E \simeq C_{\rm v, liq} \delta T_{\rm liq}$,
where $\delta T_{\rm liq} = T_{\rm liq} - T_{\rm liq}^0$,
and $C_{\rm v, liq}$ is the isochoric specific heat of the liquid.
From eqns~(\ref{eqn:alpha_liq_sol}) and (\ref{eqn:S_liq_sol}), we then have:
\begin{equation}
\alpha_{\rm liq} =
\frac{1}{1 + \exp ( - \delta T_{\rm liq} / T_{\rm int} )} \; ,
\label{eqn:alpha_liq_Fermi_fn}
\end{equation}
where the temperature interval $T_{\rm int}$ over which the
transition occurs is:
\begin{equation}
\frac{1}{T_{\rm int}} = ( N C_{\rm v, liq} / k_{\rm B} )
\left(
\frac{1}{T_{\rm liq}^0} - 
\frac{1}{T_{\rm sol}^0} 
\right) \; .
\label{eqn:melt_T_int}
\end{equation}

The Fermi function of eqn~(\ref{eqn:alpha_liq_Fermi_fn}) 
is the function we have used to
fit our simulation results for $\alpha_{\rm liq}$ as a function of
$T_{\rm liq}$ (Fig.~\ref{fig:frac_liq_96}), and the 
parameters that emerge from this
fit are $T_{\rm liq}^0 = 6705$~K, $T_{\rm int} = 124$~K.
We can now check that this value of $T_{\rm int}$ obtained by
empirical fitting is indeed consistent with 
the prediction of eqn~(\ref{eqn:melt_T_int}).
From our EAM simulations of the liquid, we obtain the
estimate $C_{\rm v, liq} / k_{\rm B} = 3.36$. Using the observed values
of $T_{\rm sol}^0$ and $T_{\rm liq}^0$, we then obtain the prediction
$T_{\rm int} = 125$~K, which is very close (perhaps fortuitously close)
to what we obtain from fitting.

It is clear from what we have said that solid-liquid alternation is
only an important effect for small systems. The key feature
that makes it easy to observe in our 96-atom system is that the
temperature distributions of the solid and liquid states
overlap significantly (Fig.~\ref{fig:hist_T_96}). 
Since the rms temperature fluctuation
of a single-phase system in the microcanonical ensemble is proportional
to $1 / \sqrt{N}$, the overlap becomes negligible for large systems.
The same conclusion is clear from 
eqns~(\ref{eqn:alpha_liq_Fermi_fn}) and (\ref{eqn:melt_T_int}), which show that
temperature interval $T_{\rm int}$ goes as $1 / N$. For a large system,
once the quasi-steady temperature $T_{\rm sol}$ of the initial solid
exceeds $T_{\rm LS}$, the homogeneous melting transition is effectively
irreversible.

Our results shed light on the Z~method for the determination of
melting properties. This method is simple to use, but our work shows
that it generally gives only an upper bound to the melting temperature
associated with a given liquid density, unless measures are taken to
correct it. This is because melting may not be observed even when the
quasi-steady temperature $T_{\rm sol}$ of the solid is above $T_{\rm
  LS}$. Indeed, melting is very unlikely to be seen if $T_{\rm sol} -
T_{\rm LS}$ is such that $\langle \tau_{\rm w} \rangle$ is much longer
than the duration of the simulation. This is a particular problem for
AIMD, where we have shown for the case of hcp Fe that, even with 50~ps
simulations on a system of 150 atoms, melting is unlikely to be seen
until $T_{\rm sol} - T_{\rm LS} \simeq 300$~K, in which case the final
liquid temperature will overestimate $T_{\rm m}$ by $\sim 300$~K. For
the systems of less than $100$ atoms and simulations of less than
$10$~ps used in some recent AIMD-Z
work~\cite{belonoshko08,belonoshko09}, the overestimation is likely be
much worse.

It is clearly desirable to have ways of correcting for the overestimate
of $T_{\rm m}$ given by the Z~method. Our work demonstrates that
calculation of the mean waiting time $\langle \tau_{\rm w} \rangle$
provides one way of doing this. For large enough $T_{\rm sol} - T_{\rm LS}$,
melting will occur rapidly, and $\langle \tau_{\rm w} \rangle$ can
then be estimated by repeating the simulation many times so as to
reduce the statistical errors on $\langle \tau_{\rm w} \rangle$.
If this is done at two or more values of $T_{\rm sol} - T_{\rm LS}$,
we then have information about the dependence of
$\langle \tau_{\rm w} \rangle^{-1/2}$ on $T_{\rm liq}$, from which the
necessary correction can be made. This would be a somewhat expensive
procedure for AIMD, but would have the great advantage of being
simple and automatic, since many simulations could be run
simultaneously on a large compute cluster. The relation with
parallel replica methods~\cite{voter98} will 
be noted. As an alternative, it may
well be possible to use Bayesian techniques to extract the
information needed for corrections from the sequences of simulations
that are required in any case by the Z~method. On a completely
different line of thought, we remark that since homogeneous
melting is a rare-event problem, it may be possible to exploit
techniques used for other rare-event problems to accelerate
melting in the Z~method. Metadynamics~\cite{laio02} might be one such technique,
since it would be easy to adopt ``collective variables'' that
would discourage the system from remaining too long 
in the solid state~\cite{angioletti10}.
We plan to investigate some of these possibilities in the future.

In conclusion, our m.d. simulations on the homogeneous melting
of the transition metal Fe confirm the existence of a rather well
defined limit of superheating, beyond which melting occurs on
a typical time-scale of ns or less. We have shown that the
statistical distribution of waiting times $\tau_{\rm w}$
before melting displays a typical ``rare event'' character,
consistent with a probability per unit time for melting to occur.
The mean waiting time $\langle \tau_{\rm w} \rangle$ lengthens
rapidly as the superheating limit is approached from above,
being roughly proportional to the inverse square of the excess
beyond the limit; it also lengthens as the system size (number of atoms $N$)
decreases, being roughly proportional to $1 / N$. This means that
the Z~method for calculating melting temperatures can be subject to
large errors if it is applied to small systems over short
times, though the method can work successfully under suitable conditions.
We have noted that these conditions have not always been satisfied
in earlier work.

\section*{Acknowledgments}

The work of DA was conducted as part of a EURYI scheme award as
provided by EPSRC (see www.esf.org/euryi). Calculations were
performed on the UCL Legion service, the HECToR national service (UK)
and used resources of the Oak Ridge Leadership Computing Facility,
located in the National Center for Computational Sciences at Oak Ridge
National Laboratory, which is supported by the Office of Science of
the Department of Energy under Contract DE-AC05-00OR22725 (USA).


\newpage

\end{document}